\documentclass[12pt,aps,prbbib,floatfix,floats]{revtex4}
\usepackage{graphicx,latexsym}
\usepackage{amsmath}
\usepackage{dcolumn}
\usepackage{bm}
\begin{document}
\title
{\bf Controlled dephasing in single-dot Aharonov-Bohm interferometers}

\author{V. Moldoveanu}
\affiliation{National Institute of Materials Physics, P.O. Box MG-7,
Bucharest-Magurele, Romania}
\author{M. \c{T}olea}
\affiliation{National Institute of Materials Physics, P.O. Box MG-7,
Bucharest-Magurele, Romania}
\author{B. Tanatar}
\affiliation{Department of Physics, Bilkent University, Bilkent, 06800
Ankara, Turkey}
\begin{abstract}

We study the Fano effect and the visibility of the Aharonov-Bohm 
oscillations for a mesoscopic interferometer with an embedded quantum 
dot in the presence of a nearby second dot. When the electron-electron 
interaction between the two dots is considered the nearby dot acts 
as a charge detector. We compute the currents through 
the interferometer and detector within the Keldysh formalism and 
the self-energy of the non-equilibrium Green functions is found up 
to the second order in the interaction strength. 
The current formula contains a correction to the 
Landauer-B\"{uttiker} formula. Its contribution to 
transport and dephasing is discussed. As the bias applied on the 
detector is increased, the amplitude of both the Fano resonance and 
Aharonov-Bohm oscillations are 
considerably reduced due to controlled dephasing. 
This result is explained by analyzing the behavior of the 
imaginary part of the self-energy as a function of energy and 
bias. We investigate as well the role of the ring-dot coupling. 
Our theoretical results are consistent to the experimental 
observation of Buks {\it et al.} [Nature {\bf 391}, 871 (1998)].

\end{abstract}

\maketitle

\section{Introduction}

The coherent nature of electronic transport through Aharonov-Bohm 
rings with embedded quantum dots (QD) was clearly established in 
various experiments \cite{e1,e2,e3,F1}. In particular the periodic 
Aharonov-Bohm oscillations of the interferometer conductance and 
the mesoscopic Fano effect were systematically 
observed and are nowadays well understood theoretically, at least 
for noninteracting dots or within mean-field 
approaches \cite{But1,HW2,Ha,AEI,EAIL,HKS,MTAT,MTGM}. 
A more subtle problem in electronic interferometry is related to 
the decoherence effects caused by inelastic processes 
like electron-electron interaction, electron-phonon or 
electron-photon coupling. More generally (see \cite{Imry,Stern}) 
decoherence appears due to the mutual interaction between a 
coherent system and its environment and 
leads to a loss of quantum interference between different electronic 
trajectories.

A particular type of decoherence is the so-called controlled 
dephasing introduced by Gurvitz \cite{Gur1} for a double dot system. 
Using one of the dots as a charge detector Gurvitz proved that 
the off-diagonal elements of the reduced density matrix 
of the measured dot are damped. Otherwise stated, the dot coherence 
is destroyed during the measurement process. Following this idea 
Buks {\it et al.} \cite{exp1} have patterned an Aharonov-Bohm 
interferometer (ABI) with a quantum point contact (QPC) located 
near the quantum dot.
The two subsystems were not coupled directly so
their mutual coupling comes only from the Coulomb interaction between
electrons in the dot and in QPC.
It was found that the transmission through the latter 
${\cal T}_{{\rm QPC}}$ increases smoothly as the plunger gate potential 
$V_g$ applied on the dot increases. 
Also, whenever a resonant conductance peak of the quantum dot is being scanned 
the QPC 'feels' the passing of a 
charge carrier through the dot and is therefore called 
"Which Path Detector" (WPD). Conversely, the current flowing 
through the WPD induces a reduction of the Aharonov-Bohm 
oscillations in the ring when the detector
is subjected to a rather large bias.
At small bias instead the visibility of the oscillations is not 
affected by dephasing. A similar experiment with coupled quantum 
dots was performed by Sprinzak {\it et al.} \cite{exp2} with a 
double quantum dot.

The first theoretical consideration of measurement dephasing in 
ABIs coupled to nearby detectors was given by 
Aleiner {\it et al.} \cite{AWM} The dephasing rate
(i.e the inverse of the time $t_d$ required for the detection of 
addition processes in the QD) was computed for weak mixing between 
the scattering states describing the WPD. 
A similar result was obtained by Levinson for a single-level 
isolated QD coupled to a conducting WPD \cite{Levi} using the 
influence functional method \cite{infl}. Another treatment 
proposed by Hackenbroich \cite{Ha} is based on the master equation 
techniques. The reduced density matrix of an isolated quantum dot 
coupled to a WPD is shown to have modified off-diagonal elements 
and the dephasing rate was computed within the Markov approximation, 
taking into account the phase change of the QPC $S$ matrix 
when one electron enters the QD. 

An alternative view on dephasing in Coulomb-coupled mesoscopic 
conductors was developed by B\"{u}ttiker {\it et al.} \cite{But2} 
The electron-electron interaction is described by geometrical 
capacitances and the dephasing rate is given in terms of the 
voltage fluctuations in QD and WPD.

In a recent work Silva and Levit \cite{SL} presented a detailed 
analysis of the dephasing rate for a quantum dot perturbed by a 
WPD by computing the interaction self-energy up to the second order 
in the Coulomb coupling constant. In the zero 
temperature limit the dependence of the dephasing rate $\nu$ on 
the ratio $eV/\Gamma$ ($eV$ being the bias 
applied to the detector and $\Gamma$ the lead-coupling ) was 
discussed.

Although the experiment of Buks {\it et al.} has triggered 
important theoretical developments the main issue of the 
contributions we just mentioned has been the calculation of 
the dephasing rate for a quantum dot which is isolated or coupled 
to two leads rather than embedded in a ring. The magnetic field 
is also absent and, to our best knowledge, no theoretical 
investigation of dephasing in the specific geometry
of Ref. \cite{exp1} was performed. Consequently an analysis of the 
Aharonov-Bohm oscillations depending on the interaction strength 
or on the WPD bias is not available. 
Moreover, the recent observation of the mesoscopic Fano effect in 
single-dot ABI \cite{F1} poses naturally the problem of possible dephasing 
effects on the Fano interference. 

In this work we consider an Aharonov-Bohm interferometer with an 
embedded dot which is coupled via a Coulomb term to a second dot 
playing the role of WPD. We take into account the geometry of the 
whole system (ring+dot+detector) and compute the currents through 
the interferometer and WPD within the Keldysh formalism. 
In particular the interferometer is a many-level system and 
therefore the current formula is not as simple as for a single-site 
dot.  Following Ref. \cite{SL}, we compute the first two 
contributions to the interaction proper self-energies.
Our calculations present for the first time clear plots showing the
effect of dephasing on the mesoscopic Fano line. We capture as well the suppresion 
of the AB oscillations when a large bias is applied on the detector and discuss
on the conditions required for the observation of the controlled dephasing. 
 
Since our main task was to discern the dephasing effects as due to the 
dot-detector interaction we neglect the intradot 
interaction and do not consider the spin degree of freedom. 
The induced decoherence due to spin-flip processes in AB 
interferometers with an interacting quantum dot without a nearby 
detector was discussed by K\"{o}nig and Gefen \cite{KG}.
Also, we do not treat in this work the dephasing effects appearing in the
 Kondo regime of the embedded dot (see \cite{Kalish} for experimental results and
\cite{SL2} for a theoretical study of a Kondo dot coupled to a nearby detector).

The outline of the paper is as follows. Section II settles 
the notation and gives the relevant formulas for currents and 
self-energies. Section III presents the numerical results and
their discussion. We conclude in Section IV. 

\section{The model and the current formula}
\subsection{The current formula}

We describe our system by tight-binding Hamiltonian and consider in a 
two-lead geometry a simple interferometer composed of three sites, 
one of them simulating the quantum dot.  
The 'Which Path Detector' is a single site coupled also to two leads
(see Fig.\,1 for geometry and notation). 
The full Hamiltonian reads:
\begin{eqnarray}\label{Hamilt}
H(t)&=&H_0+\chi_{\eta}(t)(H_{i}+H_{t}),\\
H_0&=&H_I+H_D+H_L
\end{eqnarray}
where the unperturbed Hamiltonian $H_0$ contains the Hamiltonians 
of the interferometer $I$, detector $D$, and leads $L$. The last two 
terms in Eq.\,(\ref{Hamilt}) describe the interferometer-detector 
interaction and their coupling to the leads. The two perturbations 
are adiabatically applied, more 
precisely the switching function $\chi_{\eta}(t)$ is defined 
such that $\chi_{\eta}(t)=e^{\eta t}$ for $t<0$ and 
$\chi_{\eta}(t)=1$ for $t>0$. $\eta$ is a small 
positive adiabatic parameter. 
The adiabatic switching of {\it both} the lead-system coupling 
and interaction eliminates the complications due to the Matsubara 
complex time contour which would otherwise appear in the 
perturbation theory for the Keldysh-Green
functions \cite{Wagner}. Physically this procedure means that 
the initial correlations are not taken into 
account \cite{HaugJauho}.   
The explicit form of the Hamiltonians is as follows:
\begin{eqnarray}\label{H_I}
H_I&=&\sum_{i=1}^3(\varepsilon_i+\delta_{i2}V_g)d_i^{\dagger}d_i
+\sum_{i\neq j,i,j=1}^{3}e^{i\varphi_{ij}}t_{ij}d_i^{\dagger}d_j,
\\\label{H_DH_i}
H_D&=&\varepsilon_4d_4^{\dagger}d_4,\quad
H_i=Ud_2^{\dagger}d_2d_4^{\dagger}d_4,\\\label{H_t}
H_t&=&t_{LI}(d_1^{\dagger}c_{0\alpha}+d_3^{\dagger}c_{0\beta})
+t_{}LD(d_4^{\dagger}c_{0\gamma}+d_4^{\dagger}c_{0\delta})+h.c.
\end{eqnarray}
$d_i,d_i^{\dagger}$ are annihilation/creation operators in 
the interferometer ($i=1,2,3$) and detector ($i=4$). Similarly 
we have on leads the pair $c_{},c^{\dagger}$. 
In Eq.\,(\ref{H_I}) $V_g$ simulates the gate voltage applied 
on the dot and the magnetic flux $\phi$ piercing the ring is 
included in the Peierls phases $\phi_{ij}$ attached the hopping 
constants $t_{ij}$. It is expressed in quantum flux units 
$\phi_0$ and specifically we have 
$\varphi_{12}=\varphi_{23}=\varphi_{31}=2\pi\phi/3\Phi_0$ and 
$\varphi_{ji}=\varphi_{ji}^*$. $H_i$ describes the Coulomb 
interaction of strength $U$ between the embedded dot and the detector. 
$H_t$ couples the interferometer and the detector to the sites 
$0\nu,\,\,\nu=\alpha..\delta$ 
of the lead $\nu$. The corresponding hopping constants are $t_{LI}$ and $t_{LD}$.  
For further purpose we underline here that $H_0$ and
$H_t$ could have been equally written in the first quantization form.    

Assuming that a steady-state regime is already achieved at time $t$, 
the non-equilibrium Green's function formalism gives the 
current through the lead $\alpha$ (see for example \cite{Bryant}) 
\begin{equation}\label{currF}
\langle J_{\alpha} \rangle = 
\frac{et_{LI}}{\pi\hbar}\int_{-\infty}^{\infty}
dE\, {\rm Re}G_{1\alpha}^<(E). 
\end{equation} 
where $G_{1\alpha}^<(E)$ is the Fourier transform of the real-time 
lesser Green function:
\begin{equation}\label{defG<}
G^<_{1\alpha}(t,t')=i\langle c_{0\alpha,H}^{\dagger}(t')d_{1,H}(t) 
\rangle .
\end{equation}
In the above equation $\langle \cdot\rangle$ denotes the 
statistical average on the fermionic Fock space 
with respect to the density matrix operator of the unperturbed 
time-independent Hamiltonian $H_0$. The operators are written in the interaction
picture w.r.t $H(t)$.
It is well known that the calculation of $G_{1\alpha}^<$ requires the 
knowledge of the Green-Keldysh function defined as:
\begin{equation}\label{keldysh1}
G_{mn}(\tau,\tau '):=-i\langle T_C ( d_{m,H}(\tau)
c_{n,H}^{\dagger}(\tau ')) 
\rangle . 
\end{equation}
where $T_C$ is the time-ordering operator along the 
Schwinger-Keldysh 'contour' 
$C=(-\infty, {\rm max} \{\tau, \tau '\}]\cup
[{\rm max} \{\tau, \tau '\},-\infty)$. 
The first step of the calculation is to express the mixed-indices 
Green function $G_{1,0\alpha}$ using its Dyson equation 
(see Ref. \cite{Jauho}). Since the lead $\alpha$ is coupled to 
the interferometer by a single site we have:
\begin{equation}\label{Dyson}
G_{1\alpha}(\tau,\tau')=t_{LI}\int_{C}d\tau_1G_{11}(\tau,\tau_1)
g_{0\alpha,0\alpha}(\tau_1,\tau'),
\end{equation}
where $g_{0\nu,0\nu}$ is the Green function of the semi-infinite 
lead $\nu=\alpha,\beta,\gamma,\delta$
at site $0$. This quantity is known (see Ref.\cite{Bryant}):
\begin{eqnarray}
g_{0\nu,0\nu}^{R}(E)=
\left\lbrace \begin{array}{ccc}
\frac{1}{2t_L^2}(E - i\sqrt{4t_L^2-E^2})\quad {\rm if} 
\quad E<|2t_L|  \\
\frac{1}{2t_L^2}(E - \sqrt{E^2-4t_L^2})\quad {\rm if} \quad E>|2t_L|
\end{array}\right. 
\end{eqnarray}
$t_L$ is the hopping constant on leads. Also one has:
\begin{equation}\label{rhoteta}
g_{0\nu,0\nu}^{<}(E)=2\pi i f^{\nu}(E)\rho (E)=
-2if^{\nu}(E){\rm Im}g_{0\nu,0\nu}^{R}(E).
\end{equation} 
Here $f^{\nu}(E)$ is the Fermi function for the lead $\nu$ and 
$\rho$ is the electronic density at the site $0\nu$ of the lead. 
Note that since we take 
the same hopping constant on leads $\rho$ does not depend on $\nu$. 
Expressing the contour integral in Eq.\,(\ref{Dyson}) via the 
Langreth rules \cite{Langreth}
and plugging its Fourier transform of into (\ref{currF}) one 
gets after simple calculations the following current formula 
(we shall omit the energy dependence when this 
cause no confusion):
\begin{equation}\label{curr}
\langle J_{\alpha} \rangle =
-\frac{et_{LI}^2}{\hbar}\int dE\,\rho
{\rm Im}\left ( 2G_{11}^Rf^{\alpha}+G_{11}^<\right ). 
\end{equation}
Note that due to the leads' density of states $\rho$ 
(see Eq.\,(\ref{rhoteta})) the integral runs only over
the continuous spectrum of the leads $[-2t_l,2t_L]$. 
The current formula Eq.\,(\ref{curr}) 
reduced the problem to the calculation of 
a matrix element of the interferometer Green function. 
This can be done using the Feynman-Dyson
expansion along the contour $C$. 
Without giving the straightforward details we state that the 
contour Green function 
which is a $4\times 4$ matrix satisfies the equation:
\begin{equation}\label{DysonC1}
G=G_{{\rm eff}}+G_{{\rm eff}}\Sigma_{i}G,
\end{equation}
where $\Sigma_{i}$ is the self-energy due to interaction and 
the effective Green function describes the {\it noninteracting}
system in the presence of the leads, i.e 
\begin{equation}
G_{{\rm eff}}=G_0+G_0\Sigma_{l}G_{{\rm eff}}.
\end{equation}
Here $G_0$ describes the noninteracting 
decoupled system ($QD+QPC$) and $\Sigma_{l}$ is the usual 
self energy of the leads:
\begin{eqnarray}
\Sigma_{l,mn}(z)=
\left\lbrace \begin{array}{ccccc}
t_L^2g_{0\nu,0\nu}(z)  \quad {\rm if}\quad m=n=1,3,4  \\
t_L^2(g_{0\gamma,0\gamma}(z)+g_{0\delta,0\delta}(z))\quad m=n=4\\
0\quad {\rm if}\quad m\neq n \quad {\rm or}\quad m=n=2
\end{array}\right. .
\end{eqnarray}

\subsection{The self-energies}

In what concerns the interaction self-energy $\Sigma_i$ we compute 
only the first two contributions 
$\Sigma_{i}^{1}$ and $\Sigma_{i}^{2}$ which can be identified 
from the diagrams in Fig.\,2. We used doubled lines
for electronic propagators since they are 'dressed' by the 
leads' self energy.
Using the Langreth rules for dealing with time-integrals and 
performing the Fourier transform one obtains for $m,n=1,2,3$
\begin{widetext}
\begin{eqnarray}\label{sig1}
\Sigma_{i,mn}^{R,1}(E)&=&-\delta_{m2}\delta_{n2}i\frac{U}{2\pi}\int dE'
G^<_{{\rm eff},44}(E'),\quad
\Sigma_{i,mn}^{<,1}(E)=0,\\\label{sig2<}
\Sigma^{<,2}_{i,mn}(E)&=&\delta_{m2}\delta_{n2}\frac{U^2}{2\pi^2}
\int dE_1dE_2G^<_{{\rm eff},22}(E-E_1)G^<_{{\rm eff},44}(E_2)
G^>_{{\rm eff},44}(E_2-E_1)\, .
\end{eqnarray}
\end{widetext}
The explicit forms for
$\Sigma_{i,44}^{R,1}$, $\Sigma_{i,44}^{R,2}$ and
$\Sigma_{i,44}^{R,<}$ are similar, the only difference being 
that $G_{{\rm eff},22}$ and
$G_{{\rm eff},44}$ have to be interchanged.
The retarded self-energy is related to the lesser and greater 
self-energies by the identity (see \cite{SL} ): 
\begin{equation}
\label{sig2R}
\Sigma^{R,2}_{i}(E)=\lim_{\epsilon\to 0+}\frac{i}{2\pi}\int dE' 
\frac{\Sigma^{>,2}_{i}(E')-\Sigma^{<,2}_{i}(E)}
{E-E'+i\epsilon}.
\end{equation}
To obtain the greater self-energy $\Sigma_{i,mn}^{>,2}$ one has 
just to interchange $>$ and $<$. The Dyson and Keldysh equations 
for the retarded and lesser Green functions read:
\begin{eqnarray}\label{G^R}
G^R&=&G^R_{{\rm eff}}+G^R_{{\rm eff}}\Sigma_{i}^RG^R,\\\label{G<1}
G^<&=&(1+G^R\Sigma_{i}^R)G^<_{{\rm eff}}
(1+\Sigma_{i}^AG^A)+G^R\Sigma_{i}^<G^A\\\label{G<eff}
G^<_{{\rm eff}}&=&G^R_{{\rm eff}}\Sigma_{l}^<G^A_{{\rm eff}}.
\end{eqnarray}
The last identity uses that $(G_0^R)^{-1}G_0^<=0$. A simple 
calculation gives alternative forms of (\ref{DysonC1}) and 
(\ref{G<1}):
\begin{eqnarray}\label{DysonC2}
G&=&G_0+G_0(\Sigma_{l}+\Sigma_{i})G,\\\label{G<}
G^<&=&G^R(\Sigma^<_{l}+\Sigma^<_{i})G^A\, .
\end{eqnarray}

In the following we shall use the perturbative results from above 
to compute the current
$\langle J_{\alpha} \rangle$ via Eq.\,(\ref{curr}). Using the identity
$G^R-G^A=2iG^R{\rm Im}\Sigma^RG^A$,
Eq.\,(\ref{G<}) 
and the explicit expression for $\Sigma_{l,11}$ and
$\Sigma_{l,33}$ a straightforward algebra
gives:
\begin{eqnarray}\nonumber
&\langle J_{\alpha} \rangle & = \frac{e t_{LI}^2}{\hbar}\int_{-2t_L}^{2t_L}dE
 (2\pi\rho^2G^R_{13}G^A_{31}(f^{\alpha}-f^{\beta}) \\ \label{currLB}
 &-&\rho G^R_{12}{\rm Im}(2\Sigma^R_{i,22}f^{\alpha}+
\Sigma^<_{i,22} )G^A_{21}):=J_1+J_2. 
\end{eqnarray}
The first term $J_1$ from Eq.\,(\ref{currLB}) is clearly a Landauer-type 
current. 
The second term in the current formula is a 
correction to the Landauer formula due to interaction and was not 
considered in previous papers \cite{AWM,SL}. Equation (\ref{currLB})
is a central result of the present work. 
In the following we 
provide further supporting arguments for its validity.

As emphasized in the seminal paper of Meir and Wingreen 
\cite{MW} the Landauer formula holds also for interacting systems 
in linear regime at zero temperature. The argument is based on the 
general result of Luttinger \cite{Luttinger} according to which 
the imaginary part of the self energy vanishes 
at the Fermi level, in all orders of the perturbation theory. 
The subtle point here is that if one of the two interacting 
subsystems (i.e the detector) is submitted to a finite bias one 
gets a nonvanishing imaginary 
part of the self energy even in the limit $T\to 0$.
The explicit calculation in Ref. \cite{SL} for the case of a 
single-site dot coupled to a quantum point contact clearly shows 
this fact. As we shall see in Section III the correction to the 
Landauer formula contributes non-trivially to transport and 
therefore should be kept in the current 
formula. As a consequence, the total current cannot be expressed 
in terms of the QD transmittance.
We point out that in Ref. \cite{SL} the starting point is a 
non-interacting formula for the QD transmittance $t_{{\rm QD}}$ which 
involves only the retarded Green function. The latter in turn 
contains the self-energies due to both e-e interactions and 
lead-dot coupling. However K\"{o}nig and Gefen \cite{KG} argued 
recently that such a formula may not properly define
the transmission amplitude of interacting quantum dots. 
In spite of the fact that even in the interacting case the 
integrand of $J_1$ resembles a transmission amplitude the presence
of the correction due to the ${\rm Im}\Sigma^R$ and 
${\rm Im}\Sigma^<$ prevents one to define safely
$t_{{\rm QD}}$. The rest of the paper will be mainly devoted to the 
analysis of $J_1$ and $J_2$.     

The current through the lead $\gamma$ is obtained in a similar way.
We do not give its expression here but we note that the first 
order self-energy felt by the detector 
$\Sigma^{R,1}_{{\rm i},44}=-\frac{i}{2\pi}U\int dEG^<_{{\rm eff},22}(E)=U\langle n_{22} 
\rangle$ where $\langle n_{22} \rangle$ is the electronic 
occupation number of the {\it noninteracting} dot.
Then clearly the WPD Green functions record the sudden change of  
$\langle n_{22} \rangle$ by one, 
corresponding to addition of one electron on the QD.

Although the self-energy expressions are not simple 
the calculations can be eventually performed since the effective 
lesser Green functions are given in (\ref{G<eff}) and the 
retarded functions can be computed since they describe 
noninteracting systems ($m,n=1,2,3$). Then clearly, 
one has to invert finite rank non-hermitian matrices and perform 
the energy integrals from 
Eqs.\,(\ref{sig1}), (\ref{sig2<}), (\ref{sig2R}) to get the 
self-energies expression. After some simple calculations the 
retarded self-energy of the dot reads:
\begin{widetext}
\begin{eqnarray}\nonumber
\Sigma^{R,2}_{i,22}(E)
=\frac{i}{2\pi}t_{LD}^4U^2\int dE'dE''{\tilde\rho}_{44}(E'')
\left (\frac{G^>_{{\rm eff},22}(-E')}{E+E'-E''+i\delta}-
\frac{G^<_{{\rm eff},22}(E')}{E-E'+E''+i\delta}  \right )
\end{eqnarray}
\end{widetext}
where the generalized spectral density:
\begin{widetext}
\begin{eqnarray}
{\tilde\rho}_{44}(E):=\sum_{\nu,\nu'=
\gamma,\delta}\int dE_1dE_2\delta(E_1-E_2-E)
\rho(E_1)\rho(E_2)f^{\nu'}(E_2)(1-f^{\nu}(E_1))
|G^R_{{\rm eff},44}(E_1)|^2
|G^R_{{\rm eff},44}(E_2)|^2 ,
\end{eqnarray}
\end{widetext}

This result is similar to Eq.\,(15) from 
Ref. \cite{SL} We have however extra terms
coming from the more complex geometry and the spectral density 
${\tilde\rho}_{44}$ is far more complicated that the one given in 
Eq.\,(16) from Ref. \cite{SL}. 
Now one should use Eq.\,(\ref{G<eff}) to express the effective
Green functions. Explicit 
expressions for the imaginary and real parts of 
$\Sigma^{R,2}_{i,22}$ are found by virtue of the principal-value 
formula. One can show with not much effort that in the zero 
temperature limit ${\rm Im}\Sigma_{i,22}^{R,2}(E)=0$ if there is 
no bias applied on the WPD (i.e. $\mu_{\delta}=\mu_{\gamma}$). 
Thus, there will be no dephasing as long 
as the detector is in equilibrium. \\
At this level of approximation the retarded Green function 
of the interferometer is known from the Dyson equation once we have 
computed all the effective Green functions and the first two contributions 
to interaction self-energy. 
It is now a usual Green function associated to a single particle 
operator 
\begin{eqnarray}
G^R_{44}(E)&=&\langle 4|(H_D-
\Pi_D(\Sigma^R_{l}+\Sigma_i^R)\Pi_D-E)^{-1} |4\rangle,\\
G^R_{mn}(E)&=&\langle m|(H_I-
\Pi_I(\Sigma^R_{l}+\Sigma_i^R)\Pi_I-E)^{-1}|n \rangle,
\end{eqnarray}
where $H_D$ and $H_I$ are used now in the first-quantization 
forms and $\Pi_D,\Pi_I$ are projection operators onto their associated
single particle Hilbert spaces. More explicitly, 
$\Pi_D=|4\rangle\langle 4|$ and
$\Pi_I=\sum_{i=1}^3|i\rangle\langle i|$. Also, the self-energies 
are now associated with the operators 
$\Sigma(E):=\sum_{i,j=1}^4\Sigma_{ij}(E)|i\rangle\langle j|$.
Then following the same steps as in \cite{MTGM,MTAT} we use the 
Feschbach formula to rewrite $G^R_{mn}$ is a more useful form:
\begin{equation}
G^R_{mn}(z)=\langle m|G^R_R(z)+
H_{RD}G^R_D(z)H_{DR}|n\rangle\, ,
\end{equation}
where $H_{RD}$ and $H_{DR}$ are transfer terms 
between the dot and the ring  
(i.e $H_{RD}=t_{12}e^{i\varphi_{12}}|1\rangle\langle 2|
+t_{32}e^{i\varphi_{32}}|3\rangle\langle 2| $) 
and $G^R_R =(H_R-\Pi_I\Sigma^R_{l}(z)
\Pi_I-z)^{-1}$ describes the ring without the
dot in the presence of the two leads $\alpha ,\beta$. 
$G_D$ is the effective Green function 
of the dot ($H_d$ is the single-particle Hamiltonian of 
the non-interacting uncoupled dot): 
\begin{equation}\label{Fesch}
G^R_D(z)=(H_d-\Pi_I\Sigma^R_{i}\Pi_I
-\Sigma_d(z)-z)^{-1}\, ,
\end{equation}
where $\Sigma_d:=H_{DR} G^R_R(z)H_{RD}$ is a non-interacting effective
self-energy due to the ring-dot coupling.
Obviously this quantity does not depend on bias and
can be computed explicitly by writing the $2\times 2$ matrix $G^R_R$.
We shall need $\Sigma_d$ in our calculations presented in
Section III.
The advantage of having $G^R$ in this form is two-fold. First, 
the interaction self-energy appears naturally only in the dot's 
Green function. Second, the resonant processes involving the dot are
collected in the second term and clearly separated from the 
non-interacting coherent background
contribution from the reference arm of the interferometer. 
 
\section{Results and discussion}

Before investigating the effect of the electron-electron 
interaction on the AB oscillations let us look at the Fano 
lineshapes of the current $J_{\alpha}$. 
Figure 3 shows the behavior of the Fano resonance when the 
bias $V$ on the detector increases, for two values of the 
interaction strength $U=0.3$ and $U=0.5$. 
The magnetic field is fixed at $\phi=0.0$. We begin with the low-temperature regime 
$kT=10^{-4}$ and a weak ring-dot coupling $t_{12}=t_{23}:=\tau=0.3$. 
The leads are instead strongly coupled to the detector and 
interferometer $t_{LI}=0.8,t_{LD}=1$. The bias across the leads $\alpha,\beta$ 
is denoted in the following by $V_0$.
There are two features to observe: (i) the amplitude of the Fano 
line decreases for larger bias and (ii) when $U=0.5$ 
the Fano line is shifted with respect to the resonance at $U=0.3$. 
The reduction of the Fano line is more pronounced for 
$U=0.5$. The inset in Fig.\,3 gives the current $J_{\gamma}$ 
through the detector as a function of the gate potential 
$V_g$ applied on the dot at interaction $U=0.5$. 
We take the on-site energies 
$\varepsilon_i=0$ for $i=1,2,3$ and $\varepsilon_4=0.1$. Also the hopping constant on 
leads $t_L=1$.
Clearly the detector feels the electrons passing trough the embedded dot. 
The height of the current step is a measure of the detector 
sensitivity to the electrons crossing the embedded dot. 
The inset shows that as the bias increases the response of the 
detector is enhanced. As we shall show below this suggests that 
the quantum interference within the interferometer is more 
affected by the electron-electron interaction at large bias. 
The current step does not follow precisely the Fano resonance 
for $U=0.5$ because in our approximation 
the first order self-energy of the detector $\Sigma^{R,1}_{i,44}$ 
is computed in terms of the occupation number of the noninteracting 
dot $\langle n_{22}\rangle$ (this quantity increases by one as the 
embedded dot accumulates one electron). For this reason the step-like behavior of 
$J_{\gamma}$ is correlated to the noninteracting Fano line
which is located to the left of the interacting one.  
A better correspondence between the detector characteristics and 
the Fano line can be achieved within a self-consistent calculation. 
Here we prefer to keep the lowest approximation and look for 
the role of the second order self-energy in transport.   
The two currents from Eq.\,(\ref{currLB}) are compared in Fig.\,4 
which is a zoom from the Fano dip shown in Fig.\,3 at $U=0.3$. The correction 
$J_2$ to the Landauer current $J_1$ takes significant values 
only around the Fano resonance. $J_1$ shows a dip structure depending very weakly 
on bias.  In contrast, $J_2$ increases with $V$ and there 
is a critical bias $V_c$ such that if $V>V_c$ then $J_2>J_1$. 
The numerical calculations show therefore that in the neighborhood 
of the Fano resonance $J_2$ cannot be neglected and also that 
this correction is responsible for the enhanced Fano dip seen in 
Fig.\,3. This means that $J_2$ diminishes the whole Fano line, 
hence it affects its visibility.  
          
The AB oscillations appear when the magnetic field varies and 
the gate potential is fixed to a value from the range where 
the Fano line develops. Figure 5 shows the reduction of the 
AB oscillation amplitude as the bias on the detector $V$ increases. 
In Fig.\,5a we plot oscillations around the Fano peak
($V_g=0.55$, $U=0.3$) from Fig.\,3.  The dephasing appears already at 
small bias $V=0.25$ and is considerably 
enhanced at large bias. At $V=1$ the AB oscillation amplitude 
is reduced roughly by $50\%$. A slight asymmetry 
with respect to $\phi=0$ is noticed. Around the Fano peak 
the dephasing mainly affects the ABO maxima and is due to the 
reduction in the Landauer like current $J_1$. Interestingly, 
the oscillations taken near the Fano dip $V_g=0.4$ (see Fig.\,5b) 
are more damaged because their minima are pushed upwards. 
This is due to the increasing contribution of $J_2$. In particular, 
around integer values of the magnetic flux the reduction of the 
maxima and minima are comparable. A similar dephasing effect is 
obtained for a gate potential associated to the middle of the 
Fano line (not shown). The above observations show that both 
currents contain the dephasing effect due to electron-electron 
interaction and suggest that in order to detect experimentally 
the dephasing in such systems one should look around the Fano 
dip because here the effect is stronger. 
The double-maxima aspect of the AB oscillations in Fig.\,5b is 
not an interaction effect and it is easy to explain. As known, 
the energy levels of the ring behave like $\sin \phi$. 
Single-maxima AB oscillations are obtained when the 
energy of the incident electron $E$ is close to a ring level 
while the double oscillations appear when $E$ crosses it twice 
in flux period (see examples in \cite{PhysicaE}).\\                \\

In the following we investigate the conditions under which 
the suppression of the AB oscillations due to the dot-detector 
interaction is discerned. To this end we compare the imaginary 
parts of the two self-energies appearing in the effective Green 
function of the embedded dot (see Eq. (\ref{Fesch})). Note that $\Sigma_d$ appears
only in the Landauer-like current and that $\Sigma^{R,1}_{i,22}$ 
is real. At low temperature and small 
bias applied on the interferometer the relevant range for this 
comparison is (see the integration limits in the current formula) 
$[\mu -\frac{eV_0}{2},\mu +\frac{eV_0}{2}]$. \\
In Fig.\,6 we give the imaginary parts of the two self-energies 
as a function of energy for different values of bias on the detector. 
The bias applied on the interferometer
is $V_0=0.1$ and the magnetic field is fixed at $\phi=0$. 
${\rm Im}\Sigma_d$ depends only on the energy of the 
electrons entering the interferometer while  
${\rm Im}\Sigma^{R,2}_{i,22}$ is very sensitive to both bias 
$V,$ and the chemical potential on leads $\mu$. 
For $\mu=0$ it nearly vanishes at energies smaller that 
$-0.5$ but increases as $E$ approaches the upper bound of 
the lead's spectrum $2t_L$ (recall that $t_L=1$). 
A first important observation is that in the integration 
domain $[-0.05,0.05]$ the imaginary parts 
of $\Sigma_d$ and $\Sigma^{R,2}_{i,22}$ are comparable. 
Secondly, remark that as the bias $V$ increases 
${\rm Im}\Sigma^{R,2}_{i,22}$ exceeds (in absolute value) 
${\rm Im}\Sigma_d$. Then looking at the AB oscillations from 
Fig.\,5 one infers that the dephasing effect appears as 
the interaction self-energy is of the same order as the effective 
self-energy due to the ring-dot coupling. 
The stronger suppression of the AB oscillation at bias $V=1$ 
is also understood. We have checked that by varying the magnetic 
flux the self-energies do not vary drastically and the above 
discussion still holds. 

Another interesting aspect showing the controlled feature of 
dephasing is revealed when the bias window of the interferometer 
is shifted by changing the equilibrium chemical
potential $\mu$ of the decoupled leads. The left curve in Fig.\,6 
gives again the function ${\rm Im}\Sigma^{R,2}_{i,22}(E)$ at large bias 
$V=1$ but now for $\mu=1$.
Thus, the integration domain is now $[0.95,1.05]$ and in this region
${\rm Im}\Sigma^{R,2}_{i,22}\ll {\rm Im}\Sigma_d$. We found in 
this case that the dephasing is more difficult to notice. 
Actually the oscillations of the Fano peak for zero
bias and $V=1$ do not differ significantly and a very small 
dephasing appears when a gate potential around the Fano dip 
is chosen. Therefore, the dephasing is controlled not only by
the bias applied on the environment but also by the properties 
of the dephased system, i.e by the behavior of the self-energy 
$\Sigma_d$ which depends in turn on the ring-dot
coupling and on energy. We stress that in Ref. \cite{SL} the 
leads' self-energy does not depend on
$E$ and the above discussion cannot be made.

Now, let us see what happens when the ring-dot coupling is varied. 
The analytical expression for $\Sigma_d$ shows that it behaves 
like $\tau^2$. Thus, by keeping the bias fixed and
 decreasing $\tau$ the imaginary part of $\Sigma_d$ diminishes. 
Consequently the ratio ${\rm Im}\Sigma^R_{i,22}/{\rm Im}\Sigma_d$ 
increases and from the above analysis one expects
further reduction of the AB oscillation. This is precisely 
demonstrated in Fig.\,7 where AB oscillations at bias $V=0.5$ 
are given for different couplings. Each oscillation
corresponds to the peak of the Fano line at the given $\tau$. 
Physically the enhanced dephasing noticed at smaller ring-dot 
couplings is understandable because by reducing $\tau$ the dwell 
time of electrons inside the dot increases and therefore 
they can be easily detected.  
We stress that in experiments the ring-dot coupling can be 
adjusted freely and therefore the dephasing in the system we 
consider should be easily seen for weak ring-dot coupling.    

Further analysis of the controlled dephasing is contained in 
Fig.\,8. It gives the ratio $R=A(V_i)/A_0$ between the amplitude 
of the AB oscillations at different biases $V_i$ 
and the zero-bias amplitude $A_0$. This quantity is somehow 
similar to the visibility of the oscillations measured in 
Ref.\cite{exp1}. At small temperatures we give the visibility 
for two gate potentials corresponding to the Fano peak (full line) 
and dip (dashed line).   
The behavior of $R$ confirms that in the presence of mutual 
Coulomb interaction between the detector and dot the coherency 
of the interferometer is reduced as the bias increases.
This was the main experimentally observed feature in the work of 
Buks {\it et al.} They have found that at small bias the 
visibility is almost constant while it decreases at higher bias 
and when the detector response is accurate. We note from Fig.\,7 
that the visibility of the oscillations taken around the 
Fano dip decrease faster than the once from the Fano peak. 
As we explained, this is due to the nonvanishing contribution 
of the second term in the current formula around the Fano dip. 
In order to look for temperature effects on dephasing we have 
considered as well higher temperatures $kT=0.03$ and $kT=0.05$. 
We have checked that the interaction self-energies
do not depend strongly on temperature (not shown). 
On the other hand, the effective self-energy 
$\Sigma_d$ does not depend on $T$. When the temperature increases 
the integral in the current formula runs over the whole range 
$[-2t_L,2t_L]$ and therefore it scans as well energies where 
${\rm Im}\Sigma^R_{i,22}\ll {\rm Im}\Sigma_d$. 
This could explain the slower decrease of 
the AB oscillation visibility noticed for $T=0.03$ and 
$T=0.06$.        

Before concluding let us comment on the relation and relevance 
of our calculations to the experiment of Buks {\it et al}.
 or to future experiments. We have used a detector which 
has a simple structure (a single level) and differs from the 
quantum point contact used in experiments. 
As a consequence we cannot compare directly our results 
to the experimental plots of Ref. \cite{exp1} We believe however 
that our simple model captures the main features of the experiment 
and can stimulate further measurements in order 
to check the controlled dephasing of the Fano effect. 
This topic was never considered.  

\section{Conclusions}

Motivated by the experimental paper of Buks {\it et al.} \cite{exp1} 
we have looked at the coherent transport properties of an 
AB interferometer with an embedded quantum dot which interacts 
with a nearby second dot. 
The Keldysh formalism gives the current through the ring-dot 
system and the detector. 
The interaction self-energy is computed perturbatively up to 
the second order in the interaction strength. Though the 
interaction effects on the quantum coherence were discussed in
several papers \cite{AWM,SL,But2,HKS,Jiang}, to our best knowledge 
this system was not considered theoretically before. 

The results we have obtained and discussed along this paper 
underline that: 
(i) even in the low-order perturbative approach taken here the 
electron-electron interaction causes controlled dephasing 
which manifests in the reduction of Fano line amplitude and 
the suppression of AB oscillations; (ii) in order to observe the dephasing 
a finite bias on the detector
is required but not sufficient; 
(iii) a complementary condition is that the imaginary part of 
the interaction self-energy should be of the same order as 
the self-energy coming from the ring-dot coupling; 
(iv) if the above conditions are met the dephasing effect is 
more pronounced for weak ring-dot couplings.

When the Keldysh formalism is employed to study transport 
through interacting systems having complex geometries a correction 
to the Landauer formula for the current
does not allow a proper connection to the scattering theory. 
We have shown that this correction cannot be neglected since it 
exceeds the main contribution near the Fano dip.
Moreover, as being entirely due to e-e interaction, it enhances 
the controlled dephasing. We hope our results will stimulate
further theoretical and experimental work in this area.       
  
\acknowledgments{This research has been supported by the 
CEEX Grant 2976/2005. V.M acknowledges the financial support 
from TUBITAK during the visit at Bilkent University.
We thank Prof. A. Aldea for useful discussions.
B.\,T. gratefully 
acknowledges support from TUBITAK and TUBA.}

\newpage
\begin{figure}
\includegraphics[scale=1]{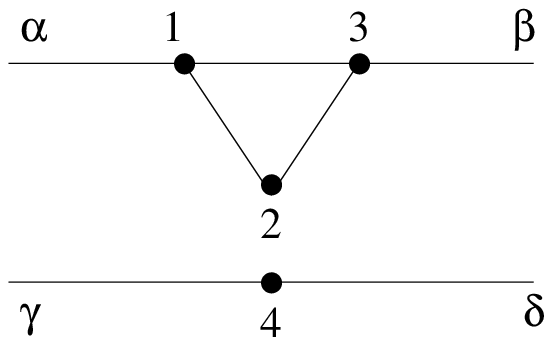}
\caption{The interferometer-detector system. The quantum dot is 
described by site $2$ which is Coulomb coupled to the single-site 
detector $4$.}
\label{figure1}
\end{figure}
\newpage
\begin{figure}
\includegraphics[scale=1]{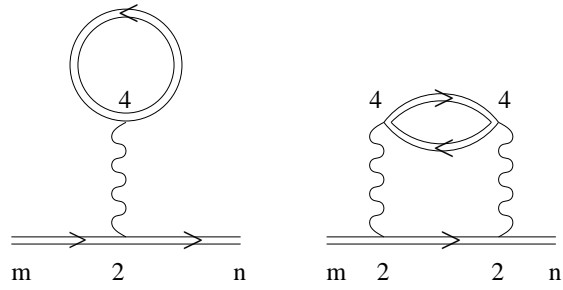}
\caption{The first two contributions to self-energy $\Sigma_i^{1}$ 
(left) and $\Sigma_i^{2}$ (right).}
\label{figure2}
\end{figure}
\newpage
\begin{figure}
\includegraphics[scale=1]{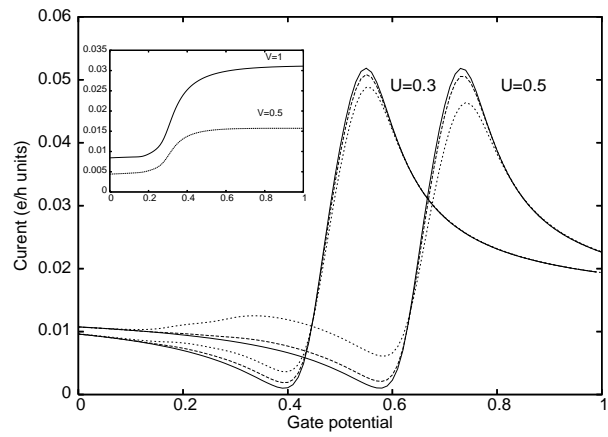}
\vskip -8cm
\caption{Coulomb-modified Fano lineshapes at different bias on 
the detector; solid line $V=0.0$, dashed-line $V=0.5$, dotted line 
$V=1.0$. Inset: the detector response at $U=0.5$.}
\label{figure3}
\end{figure}
\newpage
\begin{figure}
\hskip -4cm
\includegraphics[angle=-90,scale=1]{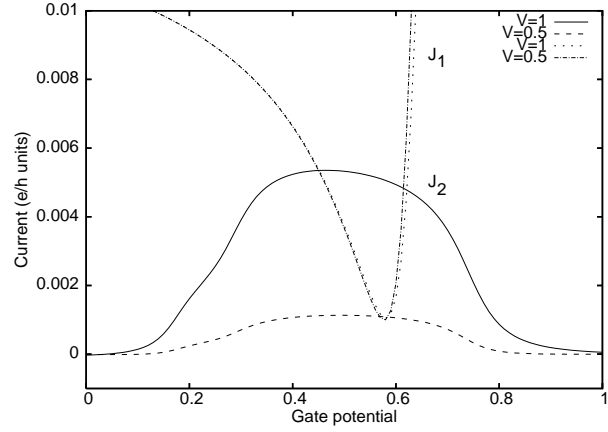}
\caption{The Landauer-like current $J_1$ (dotted and dash-dotted) 
and the correction $J_2$ (solid and long-dashed line) around the 
Fano dip. $J_1$ changes negligibly with increasing the bias $V$ 
while $J_2$ increases strongly and
even exceeds $J_1$ at $V=1.0$.}
\label{figure4}
\end{figure}
\newpage
\begin{figure}
\includegraphics[scale=1]{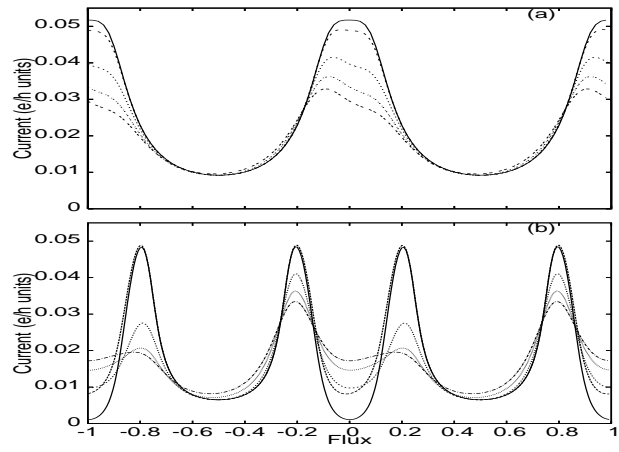}
\vskip -8cm
\caption{The reduction of the AB oscillations with increasing 
bias on the detector for two values of
the gate potential a) $V_g=0.55$ b) $V_g=0.40$. From top to 
bottom the biases are $V=0.00$, $V=0.25$,
$V=0.50$,$V=0.75$,$V=1.00$. }
\label{figure5}
\end{figure}
\newpage
\begin{figure}
\includegraphics[angle=-90,scale=1]{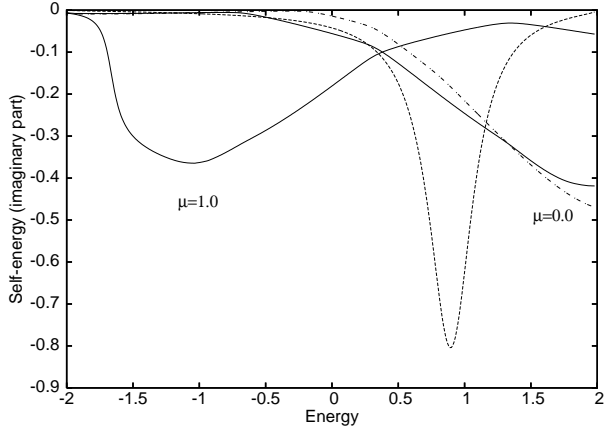}
\caption{The imaginary part of the self-energies 
$\Sigma_{i,22}^{R,2}$ and $\Sigma_d$ as a function 
of energy $E$ for fixed bias and $\mu$. ${\rm Im}\Sigma_d$ 
is plotted with dashed line and depends only
on energy. At $\mu=0.0$ two cases are shown for 
${\rm Im}\Sigma_{i,22}^{R,2}$: $V=1.0$ - full line and 
$V=0.5$ dashed - dotted.}
\label{figure6}
\end{figure}
\newpage
\begin{figure}
\includegraphics[angle=-90]{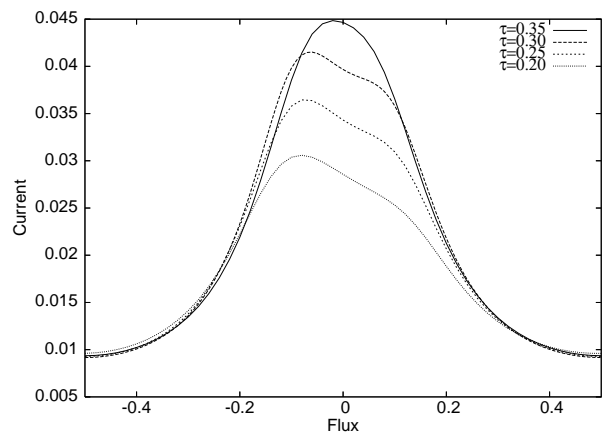}
\caption{The effect of the ring-dot coupling on the dephasing. 
As $\tau$ decreases the AB oscillations are more damaged.}
\label{figure7}
\end{figure}
\newpage
\begin{figure}
\begin{center}
\includegraphics[angle=-90,scale=1.0]{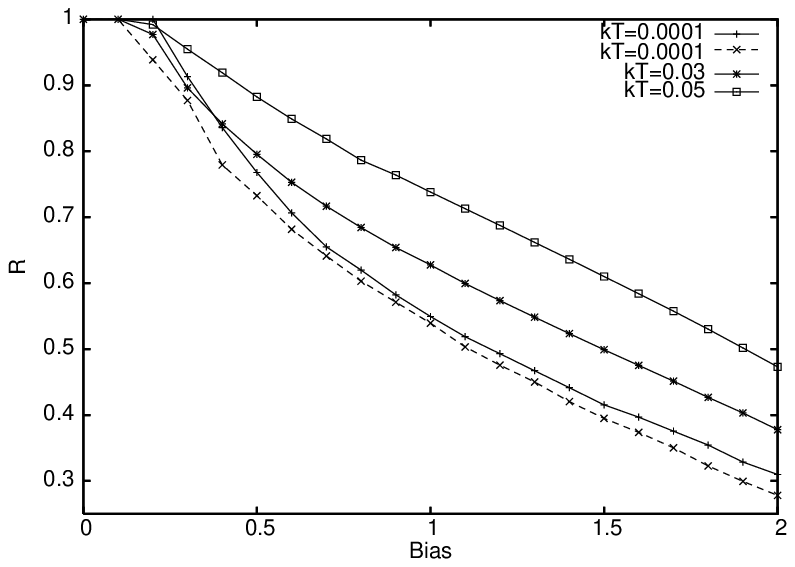}
\caption{The ratio $R$ defined in the text as a function of bias
for different temperatures. The full lines correspond to AB 
oscillations of the Fano peak while the dashed line at
$kT=0.0001$ is assigned for the Fano dip. The dephasing reduces 
as the temperature increases.}
\end{center}
\label{figure8}
\end{figure}


\begin{thebibliography}{10}

\bibitem{e1}
A. Yacoby, M. Heiblum, D. Mahalu, and Hadas Shtrikman,
Phys. Rev. Lett. {\bf 74}, 4047 (1995).
\bibitem{e2}
R. Schuster, E. Bucks, M. Heiblum, D. Mahalu, V. Umanski, and
Hadas Shtrikman, Nature {\bf 385} 417 (1997).
\bibitem{e3}
A. W. Holleitner, C. R. Decker, H. Qin, K. Eberl, and R. H. Blick
        Phys. Rev. Lett. {\bf 87}, 256802 (2001)
\bibitem{F1}
Kensuke Kobayashi, Hisashi Aikawa, Shingo Katsumoto, and Yasuhiro Iye,
Phys. Rev. Lett. {\bf 88}, 256806 (2002);
Phys. Rev. B {\bf 68}, 235304 (2003).
\bibitem{But1}
A. L. Yeyati and M. B\"{u}ttiker
Phys. Rev. B {\bf 52}, R14360 (1995).
\bibitem{HW2}
G. Hackenbroich and H. A. Weidenm{\"u}ller,
Phys. Rev. B {\bf 53}, 16379 (1996).
\bibitem{Ha}
G. Hackenbroich, Physics Reports {\bf 343}, 463 (2001).
\bibitem{AEI}
A. Aharony, O. Entin-Wohlman and Y. Imry, Tr. J. Phys. {\bf 27},
299 (2003).
\bibitem{EAIL}
O. Entin-Wohlman, A. Aharony, Y. Imry and Y. Levinson, J. Low Temp. Phys.
{\bf 126}, 1251 (2002).
\bibitem{HKS}
    W. Hofstetter, J. K\"{o}nig, and H. Schoeller,  Phys. Rev. Lett. {\bf 87}, 156803 (2001).
\bibitem{MTAT}
 V. Moldoveanu, M. \c{T}olea, A. Aldea, B. Tanatar,
 Phys. Rev. B {\bf 71} 125338 (2005).
\bibitem{MTGM}
V. Moldoveanu, M. \c{T}olea, V. Gudmundsson, A. Manolescu, Phys. Rev. B {\bf 72}, 085338 (2005).
\bibitem{Imry}
J. Imry, Introduction to Mesoscopic Physics, Oxford University Press (1997).
\bibitem{Stern}
A. Stern, Y. Aharonov, Y. Imry, Phys. Rev. A {\bf 41}, 3436 (1990).
\bibitem{Gur1}
 S.A. Gurvitz, quant-ph/9607029, Phys. Rev. B {\bf 57}, 6602 (1998).
\bibitem{exp1}
E. Buks, R. Schuster, M. Heiblum, D. Mahalu, and V. Umansky, Nature (London) {\bf 391}
, 871 (1998).
\bibitem{exp2}
D. Sprinzak, E. Buks, M. Heiblum, and H. Shtrikman,
Phys. Rev. Lett. {\bf 84}, 5820 (2000).
\bibitem{AWM}
I. L. Aleiner, N. S. Wingreen, and Y. Meir, Phys. Rev. Lett. {\bf 79}, 3740 (1997).
\bibitem{Levi}
 Y. Levinson, Europhys. Lett. 39, 299 (1997).
\bibitem{infl}
R. P. Feynman and F. L . Vernon, Ann. Phys. (N.Y.) {\bf 24} 118 (1963).
\bibitem{But2}
M. B\"{u}ttiker and A. M. Martin, Phys. Rev. B {\bf 61}, 2737 (2000).
\bibitem{SL}
A. Silva and S. Levit
Phys. Rev. B {\bf 63}, 201309 (2001).
\bibitem{Wagner}
M. Wagner, Phys. Rev. B {44}, 6104 (1991).
\bibitem{HaugJauho}
H. Haug, A.-P. Jauho,  Quantum Kinetics in Transport and Optics of
Semiconductors (Springer, Berlin,. 1996).
\bibitem{Bryant}
L. E. Henrickson, A. J. Glick, G. W. Bryant, and D. F. Barbe, Phys. Rev. B {\bf 50}, 4482(1994).
\bibitem{Langreth}
D. C. Langreth,
in  Linear and Nonlinear Electron Transport in Solids, edited by
J. T. Devreese and V. E. van Doren, Plenum Press, New York and London, 1976.
\bibitem{Jauho}
A.-P. Jauho, N. S. Wingreen, and Y. Meir, Phys. Rev. B {\bf 50}, 5528 (1994).
\bibitem{MW}
Y. Meir and N. S. Wingreen, Phys. Rev. Lett. {\bf 68}, 2512 (1992).
\bibitem{Luttinger}
J. M. Luttinger, Phys. Rev. {\bf 121}, 942 (1961).
\bibitem{KG}
J. K\"{o}nig and Y. Gefen, Phys. Rev. Lett. {\bf 86}, 3855 (2001).
\bibitem{Kalish}
M. Avinun-Kalish, M. Heiblum, A. Silva, D. Mahalu, and V. Umansky,
Phys. Rev. Lett. {\bf 92}, 156801 (2004).
\bibitem{SL2}
A. Silva and S. Levit, Europhysics Letters {\bf 62} 103 (2003).
\bibitem{Jiang}
Zhao-tan Jiang, Qing-feng Sun, X. C. Xie, and Yupeng Wang, 
Phys. Rev. Lett. {\bf 93}, 076802 (2004). 
\bibitem{PhysicaE}
A.Aldea, M.\c{T}olea, J.Zittartz, Physica E 28, 191 (2005).

\end{thebibliography}
\end{document}